\documentclass[aps,12pt]{article} 

\usepackage{epsfig,amsmath,amssymb,graphics}
\parskip=\medskipamount

\def\be{\begin{equation}}
\def\ee{\end{equation}}

\def\hfl#1#2{\smash{\mathop{\hbox to
10mm{\rightarrowfill}}\limits^{\scriptstyle#1}_{\scriptstyle#2}}}

\begin{document}
\title{\Large On the Reactions $A+A+...+A\to 0$ at a One-Dimensional Periodic 
Lattice of Catalytic Centers: Exact Solution}
\author{A.A.Naidenov$^{1}$ and S.K.Nechaev$^{2,1}$ \medskip \\
\normalsize \sl $^{1}$L.D.Landau Institute for Theoretical Physics RAN, \\ 
\normalsize \sl Kosygin str. 2, 117334  Moscow, Russia \medskip \\ 
\normalsize \sl $^{2}$Laboratoire de Physique Th\'eorique et Mod\'eles Statistiques, \\ 
\normalsize \sl Universit\'e Paris Sud, 91405 Orsay Cedex, France}
\date{\small (Published in JETP Letters, {\bf 76} (2002), 61--65)}
\maketitle

\begin{abstract}
The kinetics of the diffusion-controlled chemical reactions $A + A + ... + A\to 
0$ that occur at catalytic centers periodically arranged along a straight 
line is considered. Modes of the behavior of reaction probability $W(t)$ were 
studied at different times and different densities of the catalyst. Within the 
Smoluchowski approximation, it was rigorously proved that at large times the 
function $W(t)$ is independent of the lattice period. This means that, in the 
given asymptotic mode, the probability of the reaction on a lattice with a 
catalyst placed in each lattice site is the same as on a lattice with a 
catalyst placed in sparse sites 
\end{abstract}

In this work, we discuss the kinetics of the polymolecular irreversible 
diffusion--controlled chemical reactions $A + A + ... + A\to 0$ that occur at 
catalytic centers arranged in a periodic lattice in space. To our knowledge, 
the problem of the decrease in the amount of the reactant with time in 
catalytic diffusion--controlled chemical reactions $A + A\to 0$ was first 
studied within the random phase approximation in \cite{1}. 

The problem considered in this article ``rests on two whales'': on the one 
hand, it has characteristic features of the dynamics of diffusion--controlled 
chemical reactions, and, on the other hand, it is typical for the adsorption 
kinetics of particles diffusing in the medium of stationary ``traps'' 
(absorbers). Before going into the description of the model under study, let us 
briefly formulate the main features of its fundamental problems: 
diffusion--controlled chemical reactions and kinetics of the adsorption of 
particles on traps. 

For several decades, bimolecular diffusion--controlled reactions $A + A\to B$ 
served as the basic model for studying the kinetics of chemical reactions on 
the microscopic scale. The main ideas of the description of chemical reactions 
in the medium of diffusing reactants were formulated by Smoluchowski in 1917 
\cite{2}. Mean field methods, which were proposed by Smoluchowski and further 
developed by his numerous disciples (see, e.g., review \cite{3}) are widely 
used up to now and yield reasonable results for problems related to the 
determination of the rate of chemical reactions and change in the concentration 
of the reactant with time. However, in the case of strong stirring, i.e., when 
the diffusion of particles is significant, mean field methods become 
inappropriate because, along with pair interactions of particles, correlations 
that occur at triple, quadruple, etc. particle collisions should be taken into 
account. In works that date back to the 1980s, fluctuation effects in the 
kinetics of chemical reactions were included \cite{4}. Later the range of 
problems under study was signifi- cantly expanded by the inclusion of problems 
of chem- ical kinetics involving macromolecules, so-called 
transport--controlled chemical reactions \cite{3}. However, the comparison of 
the results that were obtained within the Smoluchowski approximation with exact 
solutions for some model systems, e.g., for a one--dimensional bimolecular 
reaction $A + A\to 0$ (without catalyst) demonstrates that the mean field 
method provides cor- rect scaling for the probability of particle decay at 
large times \cite{5,6}, which differs from the exact solution only by a 
numerical coefficient. 

The problems of determining the evolution of survival or decay probability for 
particles diffusing in $d$--dimensional space in the presence of stationary 
randomly arranged absorber traps is the subject of many studies in both the 
physicochemical and mathematical literature that have appeared over the last 
two decades or more. The great interest in this topic is obviously caused by 
its significance in studying such physical chemistry problems as 
photoconductivity and photo-- synthesis, the kinetics of binding biopolymers 
with ligands, and the adsorption of polymer molecules on surfaces with 
chemically active regions and on colloidal particles in solutions and gels. 
Works dating back to the 1970s \cite{7} demonstrated that the probability of 
the survival of independent diffusing particles in a medium with absorbing 
traps at large times cannot be described within the Smoluchowski mean field 
approximation and is controlled by the diffusion time of a particle in a cavity 
of typical size free from traps. The method that provides correct asymptotic 
relationships is identical to the optimal fluctuation method by I.M. Lifshitz 
for estimating the density of states of a disordered system in the vicinity of 
the band edge (so-called Lifshitz singularities). Thus, works \cite{7,3,4} 
stimulated the interest of mathematicians in the problem of particle decay in a 
medium of randomly arranged absorbers. Subsequent studies of model systems 
demonstrated the relation of this problem to percolation, the statistics of 
polymers in a medium of random obstacles, and some aspects of supersymmetric 
quantum mechanics \cite{4,8,9,10}. 

Now, with the general notion as to what type of problem of statistical physics 
our problem belongs, we can turn to the discussion of the model, which, as 
noted above, was formulated in \cite{1}. Let us consider a region 
(``reservoir'') in $d$--dimensional space containing an ensemble of identical 
particles A of finite size involved in Brownian motion and some quantity of 
stationary catalytic centers (traps). In the course of a random walk, there is 
the probability of encountering of $m$ particles. If this encounter takes place 
at a trap, the elementary event of the chemical reaction 
$$
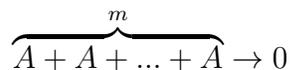
\overbrace{A+A+...+A}^{m}\to 0
$$
occurs and all the particles decay, i.e., are removed from the reservoir; if 
particles encounter outside the trap, the reaction does not occur. The task 
consists in calculating the probability $W(t|m$) of the particle decay within 
the given time interval $t$ depending on the spatial arrangement of traps. 

The probability of particle decay $W(t|m=2)$ for a bimolecular chemical 
reaction was estimated in \cite{1} within the Smoluchowski mean field theory 
for the periodic and random arrangements of traps, and the effect of catalytic 
centers was included in \cite{1} within the random phase approximation. In our 
work, we restrict our consideration to the case of $d = 1$ and the periodic 
lattice of traps. For this model, the distribution function $W(t|m)$ can be 
determined exactly without any additional assumptions. We emphasize that by the 
exact solution we mean only the possibility of the rigorous inclusion of the 
effect of the spatially distributed catalyst on the probability of particle 
decay $W(t|m=2)$. Note that, from the standpoint of real physical problem when 
a finite concentration c of a reacting particles occurs in the system, we still 
remain within the Smoluchowski approximation. At the end of the article, we 
discuss the range of applicability of the given mean field approach. Although 
one-dimensional, the problem in hand has a particular physicochemical 
application. Chemical reactions that occur at cathodes in the presence of a 
catalyst (platinum particles) can be readily described within the model under 
consideration \cite{11,12}. This is related to the fact that on the microscopic 
scale the cathode is a porous structure with one--dimensional channels in which 
the catalyst is deposited and the diffusion--controlled chemical reaction 
occurs. A problem of fundamental importance consists in the optimization of the 
adsorption process: increasing the rate of the chemical reaction and decreasing 
the concentration of the expensive catalyst. Let $L$ be the distance between 
the adjacent traps in a straight line. We denote the time dependence of the 
coordinates of the particles under consideration by $x_ 1(t), x_2(t), Е, 
x_m(t)$. The condition that all particles at the instant of time t occur at an 
arbitrary trap with the coordinate $nL$ is as follows: $x_1(t) = x_2(t) = Е = 
x_m(t) = nL$ . It is easy to realize that the random walk of m independent 
particles is equivalent to the effective one-particle random walk in the m 
-dimensional Euclidean space ${\cal E}_m = (x_1, x_2, Е, x_m)$. Thus, the 
probability $W(t|m)$ of the decay of m identical independent particles randomly 
wandering along a straight line at their simultaneous encounter at any of the 
traps is equivalent to the probability of the first occurrence of the random 
walk in the ${\cal E}_m$ space at the instance of time $t$ in an arbitrary trap 
located in the straight line defined by the equation $x_1 = x_2 = Е = x_m$. In 
our case, this statement of the problem is a starting point. The recurrence 
equation for the probability $W_t({\bf x})\equiv W(t|m,{\bf x})$ to find the 
random walk over lattice edges in the $m$--dimensional space ${\cal E}_m$ in 
the point ${\bf x}$ at the instance of time $t$ has the form 
\begin{equation} \label{eq:1}
W_{t+1}({\bf x})=D\sum_{\bf u}W_{t}({\bf x}+{\bf u})
\eta({\bf x}+{\bf u}),
\end{equation}
where $D=\frac{1}{2m}$ is the diffusion coefficient in $m$--dimensional
space, summation is performed over the nearest neighbors, and $\eta(Е)$ is the 
discrete $\delta$--function, which is defined as follows:
\begin{equation} \label{eq:2}
\eta({\bf x})=\left\{\begin{array}{cl}
0, & \mbox{if ${\bf x}$ coincides with the trap}; \\
1, & \mbox{in all the other cases}.
\end{array}\right.
\end{equation}

As usual in translation--invariant problems, it is convenient
to transform to the momentum Fourier representation:
\begin{equation}\label{eq:3}
W_t({\bf k})=\frac{1}{N^{m}}\sum_{{\bf x}}W_t(x)e^{i{\bf k}{\bf x}},
\end{equation}
where
\begin{equation}\label{eq:4}
W_t({\bf x})=\sum_{|k_1,...,k_m|<\pi}W_t({\bf k})e^{-i{\bf k}{\bf x}},
\quad k_j=\pm\frac{2 \pi s_j}{N}.
\end{equation}
As a result of the Fourier transform, Eq.(\ref{eq:1}) yields
\begin{equation}\label{eq:5}
W_{t+1}({\bf k})=\frac{1}{2m}\sum_{\bf u} e^{-i{\bf k}{\bf u}}
\sum_{{\bf q}}W_t({\bf q})\eta({\bf k}-{\bf q}),
\end{equation}
where
\begin{equation}\label{eq:6}
\eta({\bf k)}=\delta({\bf k})-\frac{1}{N^{m}}\sum_{|nL|<N/2}
e^{i k_1 n L}.
\end{equation}
In Eq.(\ref{eq:6}), we assumed that the straight line in which the traps are 
located coincides with the $[0, x_1)$ axis in the ${\cal E}_m$ space. Let us 
use the equation
$$
\sum_{n=-\frac{N}{2L}}^{\frac{N}{2L}}e^{i k_1 nL}=\frac{N}{L}
\sum_{|n|<\frac{L}{2}}\delta\left(k_1-\frac{2\pi n}{L}\right),
$$
in which, for the sake of definiteness, we assume $\left[\frac{N}{2L}\right]\le=
\frac{N}{2L}$. Using the latter expression, Eq.(\ref{eq:5}) can be written in the 
form
\begin{multline}\label{eq:7}
W_{t+1}(k)=\frac{1}{m}(\cos k_1+...+\cos k_m)\Bigg\{W_{t}(k)-
\frac{1}{N^{m-1}L}\sum_{q_2,...,q_m \atop |n|<L/2}
W_t\left(k_1-\frac{2\pi n}{L},q_2,...,q_m\right)\Bigg\}.
\end{multline}
where 
$$
-\pi\le \{k_1,...,k_m\}<\pi, \qquad-\pi\le \{q_2,...,q_m\}<\pi
$$
Equations of this type are conveniently solved by the generating function 
method: 
\begin{equation}\label{eq:laplace}
W({\bf k},s)=\sum_{t=0}^{\infty} W_t({\bf k}) s^t, \quad
W(t)=\frac{1}{2\pi i}\int\limits_C \frac{W(s)ds}{s^{t+1}}.
\end{equation}
Multiplying both parts of expression (\ref{eq:7}) by $s^t$, after simple algebraic
transformations we obtain
\begin{equation}\label{eq:W(k,s)}
W({\bf k},s)=\frac{W_0({\bf k})}{1-A({\bf k},s)}-
\frac{A({\bf k},s)}{1-A({\bf k},s)}
\frac{S\Big(\frac{W_0({\bf k})}{1-A({\bf k},s)}\Big)}
{S\Big(\frac{1}{1-A({\bf k},s)}\Big)},
\end{equation}
where the following designations are introduced:
\begin{equation}\label{eq:A}
A({\bf k},s)=\frac{s}{m}(\cos k_1+...+\cos k_m),
\end{equation}
$$
S\big(W({\bf q})\big)=
\frac{1}{N^{m-1}L}\sum_{q_2,\dots,q_m \atop |n|<L/2}
W\Bigg(q_1-\frac{2\pi n}{L},q_2,\dots,q_m\Bigg).
$$

Let us select uniform starting conditions, i.e., $W_0({\bf k})=
\delta(k_1)\dots\delta(k_m)$ and proceed to limit $N\to\infty$:
\begin{multline} \label{eq:W(k,s)gen}
W({\bf k},s) = \frac{\delta({\bf k})}{1-A({\bf k},s)}-
\frac{A({\bf k},s)}{1-A({\bf k},s)}\frac{1}{(1-s)} \times
\left(\sum_{q_1}\int\cdots \int \frac{dq_2
\ldots dq_m}{1-\frac{s}{m}(\cos q_1+\ldots+\cos q_m)}
\right)^{-1},
\end{multline}
where $q_1$ takes on the values $q_1=k_1-\frac{2\pi n}{L},\;|n|<L/2$. The 
second term, which enters into the expression with the minus sign, 
describes particle decay at traps. We are interested in the zeroth 
harmonic of this term, which determines the probability of the reaction
\begin{equation}\label{eq:Linfsum}
W(s)=\frac{s}{(1-s)^{2}}\times
\left(\sum_{q_1}\int\cdots \int\frac{dq_2\ldots dq_m}
{1-\frac{s}{m}(\cos q_1+\ldots+\cos q_m)}\right)^{-1}.
\end{equation}

For further consideration of the problem, we must evaluate the integral
\begin{equation}\label{eq:defI}
I(\alpha)=\int\cdots\int\frac{dq_2\ldots dq_m}
{\alpha-(\cos q_2+\ldots+\cos q_m)},
\end{equation}
where $\alpha=\frac{m}{s}-\cos q_1$. This function can have poles only on 
two points $\alpha=\pm (m-1)$, i.e. $s_{1,2}=\frac{m}{\cos q_1 \pm 
(m-1)},\quad s_{1,2}\ge 1$. Integral (\ref{eq:defI}) in these regions is 
determined by the values of $q_i$, which are nearly zero; therefore, $\cos(Е)$ 
in the integrand can be expanded. The change $\cos q_i\approx 
1-\frac{1}{2}q_i^2$ corresponds to the changing from the lattice to the 
continuous limit with respect to the $i$th coordinate. In the vicinity of the 
point $\alpha=m-1$, $\epsilon=\alpha-m+1$ we have
\begin{equation}
I_m(\epsilon)=2\int \cdots\int \frac{dq_2 \ldots dq_m}
{2\epsilon + q_2^2+\ldots+q_m^2}.
\end{equation}
Changing to spherical coordinates ($S_{mЦ1}$ is the area of the sphere in the 
$(mЦ1)$--dimensional space, $A$ is some constant, $A \sim р$), we obtain
\begin{equation}
I_m(\epsilon)\approx 2S_{m-1}\int_0^A \frac{q^{m-2}dq}{2\epsilon+q^2}.
\end{equation}
The values of $I_m(\epsilon)$ at different values of $m$ are
\begin{equation}\label{Im_approx}
I_m(\epsilon) = \left\{
\begin{array}{*{20}cl}
\frac{2\pi^2}{\sqrt{2\epsilon}}, & \mbox{при $m = 2$} \medskip; \\
8\pi\log\left(1+\frac{A^2}{2\epsilon}\right), & \mbox{при $m=3$} \medskip ;\\
\sim A^{m-3}, & \mbox{при $m\ge 4$}.
\end{array} \right.
\end{equation}

The probability of decay at traps within these designations
is
\begin{equation}\label{eq:WI}
W(s) = \frac{m}{L(1-s)^2 \frac{1}{L}\sum\limits_{p=-L/2}^{L/2}
I\left(\frac{m}{s}-\cos \frac{2\pi p}{L}\right)}.
\end{equation}
The integrand in the function $W(t)$ has singularities $s^t$ and $(1-s)^2$. We 
are interested in the behavior of $W(t)$ governed by the vicinity of the point 
$s=1$, which corresponds to$\alpha=m-\cos q_1$. Hence, the expression for the 
generating function $W(s)$ within a numerical multiplier can be rewritten as 
follows: 
\begin{equation}\label{Wprecise}
W(s)\sim \frac{1}{(1-s)^2 \sum\limits_{p=-L/2}^{L/2}
I_m\left(\frac{m}{s}-\cos\frac{2\pi p}{L}-m+1\right)}.
\end{equation}

The time dependence $W(t)$ is restored from the generating function $W(s)$ by 
inverse Laplace transform (\ref{eq:laplace}). The contour integral depends on 
singular points of the integrand, and the pole at the point $s = 1$ makes the 
largest contribution. Recall that the probability $W(t)$ of particle decay 
unambiguously characterizes the decrease in the reactant concentration with 
time and, thus, determines the effective rate constant of the chemical 
reaction. Further calculations will be based on the Tauberian theorem, which 
readily provides asymptotic estimations in the cases of interest without 
explicit use of the inverse Laplace transform. 
\medskip

\noindent {\bf Reactions with $m = 1$ and $m\ge 4$}. Retaining only the
divergent part in expression(\ref{eq:Linfsum}), we obtain the asymptotics
\begin{equation}
W_{m=1}(t\to\infty)\sim\sqrt t.
\end{equation}

In the case of $m\ge 4$ the result is also readily obtained
because $I_m\sim {\rm const}$ in the region of $s=1$. Therefore, the
behavior of the function at large times is governed by
the pole $\frac{1}{(1-s)^2}$, which corresponds to
\begin{equation}\label{eq:m>3}
W_{m\ge 4}(t\to\infty)\sim t.
\end{equation}

\noindent {\bf Reactions with m = 2}. The behavior of $W(t)$ at $m = 2$
is more interesting. In this case, the generating function
$W(s)$ of the absorption probability is as follows:
\begin{equation}\label{W2precise}
W(s) \sim \frac{1}{(1-s)^2\sum\limits_{p=-L/2}^{L/2} \left(2-
s\cos\frac{2\pi p}{L}-s\right)^{-1/2}}.
\end{equation}
Applying the Tauberian theorem and replacing the sum with the integral, 
we obtain
\begin{equation}
W_{m=2}(t)\sim \frac{t}{\sqrt{t}+\sigma^{-1}+\frac{2L}{\pi}
\left[\log{4}-\log\left(\frac{\pi}{L}+\sigma\right)\right]},
\end{equation}
where $\sigma=\sqrt{\frac{\pi^2}{L^2}+\frac{1}{t}}$. In the two limiting cases
\begin{equation}
W_{m=2}(t)\sim \left\{
\begin{array}{ll}
\frac{t}{t^{1/2}+{\rm const}} \sim \sqrt{t},&t\gg \frac{L^2}{\pi^2}\label{eq:m*}\\
\frac{t}{L\log8t},&1\ll t\ll \frac{L^2}{\pi^2}.
\end{array}
\right.
\end{equation}

\noindent {\bf Reactions with m = 3}. Expression (\ref{Im_approx}) for $I_{m=3}$
gives the correct relationship; however, it contains an
undefined constant, which complicates estimations. By
the integration of Eq. (\ref{eq:defI}) without expanding cosines in
series, we can obtain the more accurate formula
\begin{equation}
I(\alpha)\approx\pi \log \left( \frac{16}{3-2s-s\cos q}\right).
\end{equation}
The asymptotics for $t\to\infty$ can be obtained directly by
applying the Tauberian theorem to generating function
(\ref{eq:WI}), which leads to the following expression:
\begin{equation} \label{eq:m=3}
W_{m=3}(t)  \approx  \frac{t}{C+\log{t}-\log\left(\frac{3}{t}+
\frac{2\pi^2}{L^2}\right)}.
\end{equation}
where $C=5 L\log{2}+4\log\left(\frac{2\pi}{L}\right)-4-\log\frac{4}{3}$.

Absorption probabilities that were obtained above
are insufficient for the exact calculation of the reaction
rate but are directly related to the Smoluchowski constant
$K_{Smol}(t)=\frac{dW(t)}{dt}$. Indeed, this constant by defi-
nition is the probability of the decay of a solitary particle
(i.e., without inclusion of cooperative effects) at the
catalyst trap. Within the Smoluchowski approximation,
the concentration of particles in the polymolecular
reaction of $m$ particles is determined from the kinetic
equation
\begin{equation}
\frac{d C(t)}{dt}=-K_{Smol}(t)C_{tr}C^m(t)
\end{equation}

After simple calculations, we obtain ($C_0=C(t=0),\;m>1$)
\begin{equation} \label{eq:c(t)}
C(t)=\left(C_0^{1-m}+(m-1)C_{tr}W(t)\right)^{-1/(m-1)}.
\end{equation}

The Smoluchowski approximation is valid in cases when stirring in the system 
due to the diffusion of particles is a slower process than the chemical 
reaction event. That is, if the motion of particles is nearly absent, the 
chemical reaction event involves only the particles that are occasionally the 
closest to each other. This is the reason why the Smoluchowski approximation is 
adequate at a low density of particles in the system, when the free diffusion 
path of particles is sufficiently large compared to the time of the chemical 
reaction event. To demonstrate the validity of the mean field approximation in 
the problem under consideration for any distance $L$ between the catalytic 
centers, it is sufficient to show that the Smoluchowski theory works in the 
two limiting cases: (a) at $L = 1$, i.e., when the reaction occurs each time on 
the encounter of a pair of particles and, consequently, is independent on the 
position of the catalytic center; and (b) at $L\to \infty$, i.e., when the system 
contains a solitary catalytic center. The exact solution of the many-particle 
problem in case (a) for bimolecular reactions was reported in \cite{5,6}. 
As noted  in the introduction, the exact and mean-field solutions have the same 
asymptotics and differ only by a numerical coefficient. For the case (b), the 
decrease in concentration $c(t)$ because of the chemical reaction can be easily 
estimated ($m\ge 2$):
\begin{equation} \label{eq:chem}
\frac{dc_{\textrm{chem}}(t)}{dt}=-K_{\textrm{chem}} 
c^m_{\textrm{chem}}(t),\quad \Rightarrow \quad
c_{\textrm{chem}}(t) \sim \frac{c_0}{K_{\textrm{chem}}}\,t^{-1/(m-1)}
\end{equation}

In order to find the decrease in concentration $c(t)_{\textrm{diff}}$ 
because of diffusion in the case of the reaction of $m$ particles
at a center, we must solve the diffusion equation
in $m$--dimensional space, where each coordinate corresponds
to the concentration of one of the particles
involved in the reaction. As can be easily seen,
\begin{equation} \label{eq:diff}
\begin{array}{ll}
\displaystyle
c_{\textrm{diff}}(t) \sim c_0\, \frac{\log t}{t} & (m=2) \\
\displaystyle
c_{\textrm{chem}}(t) \sim c_0\, {\rm const} & (m=3,4,...)
\end{array}
\end{equation}

For all $m\ge 2$, the following condition is fulfilled: after some instant of 
time $t = t(c_0, K_{chem})$, the concentration because of diffusion is larger 
than the concentration because of the chemical reaction, and, consequently, the 
rate of diffusion is lower than the rate of the chemical reaction, which means 
that the Smoluchowski approximation is valid. 

Of particular interest are the relationships $W_m(t)$ at large times (при 
$t\to\infty$), namely, expressions (\ref{eq:m>3}), (\ref{eq:m*}) and 
(\ref{eq:m=3}). As can be seen from the corresponding formulas, the probability 
of reaction $W_m(t)$ is independent of the period of the lattice of catalytic 
centers. This means that in the given asymptotic mode the probability of the 
reaction on the lattice with the catalyst placed in each lattice site is the 
same as on the lattice with the catalyst placed in sparse sites. Recall that 
this result was first formulated in \cite{1} within the random phase 
approximation. Thus, our work can be considered as rigorous proof of the 
effect, which has promising technical applications. We are grateful to G. 
Oshanin for helpful discussions of the work. The work was supported by the 
Russian Foundation for Basic Research, project No. 00--15--99302.

\end{document}